\documentclass[11pt]{JHEP3}

\usepackage{cite,slashed}
\usepackage{amsmath,amstext,amssymb}
\usepackage{psfrag}
\usepackage[dvips]{graphicx}  
\usepackage{amsmath,amsfonts}
\usepackage{bm}
\newcommand{\romanNum}[1]{\@roman{#1}}

\def\be{\begin{eqnarray}}
\def\ee{\end{eqnarray}}
\def\beq{\begin{equation}}
\def\eeq{\end{equation}}

\def\({\left (}
\def\){\right )}
\def\[{\left [}
\def\[{\right ]}

\def\bc{\begin{center}}

\def\ec{\end{center}}
\def\be{\begin{eqnarray}}
\def\ee{\end{eqnarray}}

\title{One dimensional $s$-wave holographic superconductor with supercurrent}
\author{
 Hua-Bi Zeng $^{a,b}$ \\

$^{a}$School of Mathematics and Physics, Bohai University\\
JinZhou 121000, China \\
$^{b}$Department of Physics, Nanjing University\\
Nanjing 210093, China \\
Email: \email{zenghbi@gmail.com}
}

\abstract{
 We study the one dimensional $s$-wave holographic
superconductor by turning on the
vector potential $A_x$ in the bulk, which behaves as
$A_x=A_x^{(0)} \ln z+ A_x^{(1)}$ on the boundary.
By solving the model with fixed  $A_x^{(0)}$, we find that if we identify the $A_x^{(0)}$
with the supercurrent $j_x$ of the holographic superconductor,
the results agree with the Ginzburg- Landau theory.
For example, $A_x^{(0)}$ will break the superconductivity, and the
critical value of $A_x^{(0)}$ is proportional to $(T_c-T)^{3/2}$.
}
\preprint{\today}


\begin{document}
\maketitle

\section{Introduction}
The AdS/CFT correspondence \cite{1,2,3,4} has been taken as a very
useful method to study strongly coupled
phenomena now. A successful application of
the AdS/CFT correspondence to condensed matter physics
is to study the superconductors.
The strategy to build a holographic superconductor
is to couple an $AdS$ black hole with some
charged field and the $U(1)$ gauge fields.
The black hole will have hair when the temperature
of the black hole is low enough.
The charged hair breaks the $U(1)$ gauge
symmetry in the bulk and the black hole
is superconducting \cite{5}.
By putting a charged scalar field coupled to the $U(1)$ gauge field in $AdS_4$ black hole, one can get the $s$-wave holographic superconductor in which the order parameter is a scalar\cite{6,7}. By putting a pure $SU(2)$ gauge field in the bulk theory, one can get the $p$-wave holographic superconductor with a vector order
parameter \cite{8,9,10,11,12}. In order to build a $d$-wave holographic
superconductor we need a charged tensor field coupled to a $U(1)$ gauge
field in the bulk that leads to a tensor order parameter \cite{13,14,15}.

Most of the studies on holographic superconductors have focused on the $(2+1)$ dimensional
systems by using AdS$_4$/CFT$_3$. The $(3+1)$ dimensional
holographic superconductor shows similar
properties as the $(2+1)$ dimensional superconductors, as
studied in\cite{16} by using AdS$_5$/CFT$_4$.
The $(1+1)$ dimensional $s$-wave and $p$-wave holographic superconductors were explored in
\cite{17} and \cite{18} respectively by using AdS$_3$/CFT$_2$.
Due to the logarithm behavior of $A_x$ on the boundary,
we have to realize that the AdS$_3$ realization of holographic superconductor
is an honest superconductor with a dynamic gauge field on the boundary,
in which the local boundary symmetry is spontaneously broken \cite{18}.
The holographic superconductors with dynamic gauge fields in higher dimensions were
studied in \cite{dynamic1,dynamic2,dynamic3,dynamic4}.
A holographic quantum liquid in (1+1) dimensions from a probe D3 brane in the AdS Schwarzschild planar black hole
background was studied in \cite{19}.

In this paper, we studied the one dimensional $s$-wave holographic superconductor by switching on the
vector potential $A_x$ in the bulk theory. $A_x$ behaves as
$A_x=A_x^{(0)} \ln z+ A_x^{(1)}$ on the AdS$_3$ boundary.
By solving the model with fixed $A_x^{(0)}$, we find that $A_x^{(0)}$
plays the role of supercurrent in a superconductor.
If we identify $A_x^{(0)}$ with the supercurrent $j_x$ and
$-A_x^{(1)}$ as the superfluid velocity $v_x$,
the results agree with the Ginzburg- Landau (G-L) theory for
a superconductor with supercurrent.
For example, $A_x^{(0)}$ will break the superconductivity, the
critical value of $A_x^{(00}$ being proportional to $(T_c-T)^{3/2}$.
These results indicate that we can interpret the source
term $A_x^{(0)}$ as the supercurrent $j_x$ while $-A_x^{(1)}$ as the
superfluid velocity $v_x$.
The G-L theory results we found in $(1+1)$ dimensional $s$-wave holographic
superconductors have also been founded in the $(2+1)$ dimensional holographic
superconductors with supercurrent\cite{20}.
In analogy,
for the scalar potential $A_t=A_t^{(0)} \ln z+A_t^{(1)}$ on the boundary,
$-A_t^{(0)}$ can be interpreted as the
charge density $\rho$ while $A_t^{(1)}$ as the chemical
potential $\mu$.

The organization of this paper is as follows. In Section
$2$ we first introduce the Einstein-Maxwell-charged
scalar system in AdS$_3$ black hole, and then confirm that there
is a continuous phase transition when $A_x=0$ by free energy calculations.
In Section $3$ we solve the system with fixed value of $A_x^{(0)}$ and we find
that $A_x^{(0)}$ should be the supercurrent of the holographic superconductor.
We also try to solve the system with fixed value of $A_x^{(1)}$ in Section $4$.
Discussions and conclusions are given in Section $5$.

\section{The condensed phase with $A_x=0$ }
In this section we firstly review the results in \cite{17}. We also compute
the free energy to confirm the existence of the continuous
superconducting phase transition.
\subsection{The Einstein-Maxwell-charged
scalar system in $AdS_3$ black hole}
The Lagrangian of the Einstein-Maxwell-charged
scalar system reads
\begin{equation}
\mathcal{L}=-\frac{1}{4}F^{ab}F_{ab}-\frac{1}{2}m^2 |\Psi|^2-|(\partial_\mu-iA_\mu)\Psi|^2.
\end{equation}
The neutral $AdS_3$ black hole background in Poincar\'{e} coordinates is
given by~\cite{17}
\begin{equation}
ds^2=\frac{L^2}{z^2}\left(-f(z)dt^2+dx^2+\frac{dz^2}{f(z)}\right) \, ,
\end{equation}
in which $f(z)=1-z^2$, $z=r_+/r$, and $r_+$ is the horizon of the black hole.
We set $L=1$ and $r_+=1$.
The black hole temperature is given by
\begin{equation}
T=\frac{r_+}{2 \pi } \, .
\end{equation}
With the ansatz
\begin{equation}
\Psi=\Psi(r), A_t=\Phi(r), A_x=A_x(r),
\end{equation}
we have the equations of motion (EOMs) for the three fields,
\begin{equation}
\Psi''-\frac{2z}{1-z^2}\Psi'-\frac{1}{z}\Psi'+\frac{\Phi^2 \Psi}{(1-z)^2}-\frac{A_x^2 \Psi}{1-z^2}-\frac{m^2\Psi}{z^2f}=0,
\end{equation}

\begin{equation}
\Phi''+\frac{1}{z}\Phi'-\frac{2\Phi \Psi^2}{z^2(1-z^2)}=0,
\end{equation}

\begin{equation}
A_x''-\frac{2z}{1-z^2}A_x'+\frac{A_x'}{z}-\frac{2A_x\Psi^2}{z^2(1-z^2)}=0.
\end{equation}

The boundary behavior for the two Maxwell fields $\Phi$ and $A_x$ at $z=0$ are
\begin{equation}
\Phi= \Phi^{(0)} \ln (\Lambda z)+\Phi^{(1)}+ \cdots,
\end{equation}

\begin{equation}
A_x=A_x^{(0)}\ln (\Lambda z)+A_x^{(1)}+ \cdots,
\end{equation}
where $\Lambda$ is the renormalization scale included in the logarithm.
We set $\Lambda$ to be $1$ without loss of generality.
It has been found that there is only one way to quantize the theory (alternative quantization)
by defining the coefficients of the logarithmic terms to be the sources \cite{21,22,23}.

\begin{figure}
\begin{center}
\includegraphics[width=10cm,clip]{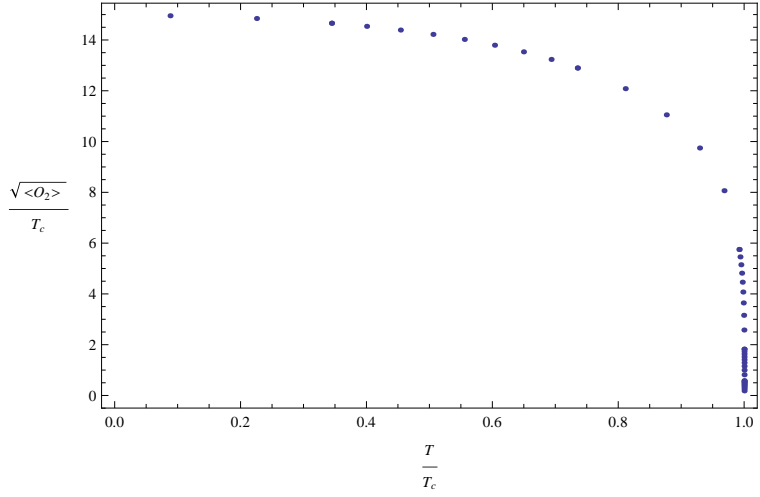}
\caption{The
condensate versus temperature when $A_x$ is zero. \label{fig:condensation} }
\end{center}
\end{figure}

As for the scalar field $\Psi$, we can only consider the case of $m^2=0$,
since the results for
other values of $m^2$ are similar \cite{17}.
The asymptotic behavior of $\Psi$ near the boundary for $m^2=0$ reads
\begin{equation}
\Psi=\tilde{\Psi}^{(1)}+ \tilde{\Psi}^{(2)} z^2+ \cdots.
\end{equation}
We can only choose the scalar operator as $ \langle O_2 \rangle=\tilde{\Psi}^{(2)}$, with the boundary
condition $\tilde{\Psi}^{(1)}=0$\cite{17}.  $ \langle O_2 \rangle$ is the order parameter of the superconductor.
On the horizon, regularity requires $\Phi=0$, with the other two fields $\Psi$ and $A_x$
being finite.
In this section we set $A_x=0$ and solve the two EOMs
Eq. (2.5) and Eq. (2.6)
for $\Phi$ and $\Psi$ with
vanishing $A_x$.

\subsection{Phase diagram and the free energy analysis }
The phase diagram of the superconductor with $A_x=0$ is plotted in Fig. 1.
It is clear that there is a continuous
phase transition at $T_c$. After computing
the free energy density we will confirm the
result further.

According to the AdS/CFT dictionary,
the free energy of the dual field theory
is related to the on-shell action of the
classical gravity theory,
\begin{equation}
\mathcal{F}=-T S_{os}+...
\end{equation}
where the ellipsis denotes boundary terms we may need.
Employing the equations of motion, the on-shell action
can be written as
\begin{equation}
 S_{\text{os}}=\int dx^2(\frac{z}{2} \Phi \Phi' -\frac{z f}{2} A_x A_x'-\frac{f}{z} \Psi \Psi')|_{z=0}+\int dx^3(\frac{A_x^2 \Psi^2}{z}-\frac{A_t^2 \Psi^2}{zf}).
\end{equation}
By plugging in the asymptotic behaviors of $\Phi$, $A_x$ and $\Psi$ we get

\begin{equation}
S_{\text{os}}=\int dx^2(\frac{1}{2} \Phi^{(0)}(\Phi^{(1)}+\Phi^{(0)} \ln z)-\frac{1}{2}A_x^{(0)}(A_x^{(1)}+A_x^{(0)} \ln z))|_{z=0}+\int dx^3(\frac{A_x^2 \Psi^2}{z}-\frac{A_t^2 \Psi^2}{zf}).
\end{equation}
In order work in the ensemble with fixed $A_i^{(0)}$, we need to add the boundary \cite{21,22}
\begin{equation}\label{eq:Sbdy}
S_{\text{bdy}} =  \frac{1}{2} \int dt\, dx\, \left(\sqrt{-h} A_\mu F^{z\mu}\right)|_{z=0}
 \, ,
\end{equation}
where $h$ is the induced metric of the boundary.
We can see $S_{\text{os}}+S_{\text{bdy}}$which is divergent. In order to cancel the divergence, we add the counter term\cite{22}
\begin{equation}
S_{\text{CT}}=\int dx^2(\sqrt{-h}F_{zt} F^{zt} \ln z+ \sqrt{-h}F_{zx} F^{zx} \ln z)|_{z=0},
\end{equation}

Finally, the free energy density for the superconductor reads
\begin{equation}
F=\frac{\mathcal{F}}{l}=- \Phi^{(0)}\Phi^{(1)}+ A_x^{(0)} A_x^{(1)}-\int dz(\frac{A_x^2 \Psi^2}{z}-\frac{A_t^2 \Psi^2}{zf}),
\end{equation}
where $l$ is the length of the superconducting wire.
The superconductor is uniform since the order
parameter is independent of $x$.

\begin{figure}
\begin{center}
\includegraphics[width=8cm,clip]{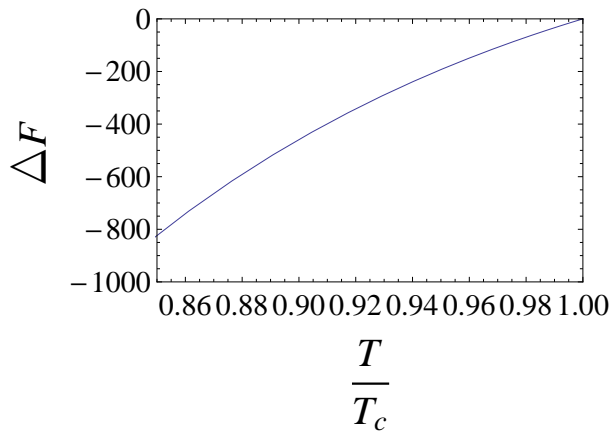}
\caption{The free energy difference between the superconducting
state and the normal state with $A_x=0$. It is
clear that the superconducting state is stabler when $T<T_c$. \label{fig:fenergy} }
\end{center}
\end{figure}

In Fig. 2 we plot the free energy density difference
between the superconducting and normal states.
It can be seen that the superconducting state
is more stable than the normal state when $T<T_c$.

\section{The superconductor with fixed $A_x^{(0)}$ }
In AdS$_4$ setup of holographic superconductor, $A_x=A_x^{(0)}+A_x^{(1)} z$
on the boundary. It is clear now
we can interpret $-A_x^{(0)}$ as the
superfluid velocity and $A_x^{(1)}$
as the supercurrent.
There are two kinds of quantizations
for $A_x$, which means that we can
fix the value of $A_x^{(0)}$ and let $A_x^{(1)}$
free to fluctuate, or fix the value of
$A_x^{(1)}$ and let $A_x^{(0)}$
free to fluctuate.
The two methods to solve the problem correspond to experiments where the current,
or the superfluid velocity is kept fixed.
It has been shown \cite{20,24,25,26} that
by fixing $A_x^{(0)}$ or $A_x^{(1)}$
on the boundary, the $AdS_4$ black hole holographic
superconductors (both $s$-wave and $p$-wave holographic superconductors) with fixed supercurrent
or fixed superfluidity velocity
have the same results as G-L
theory.

However, since the gauge fields
behave much differently on AdS$_3$
boundary,
it is very interesting to ask
what will happen when the
superconducting black hole has
non-vanishing vector potential $A_x$.
In this section we solve the EOMs with fixed value of $A_x^{(0)}$ in Eq. (2.9).
In the next section we also try to solve the EOMs with fixed value
of $A_x^{(1)}$ in Eq. (2.9).

\subsection{The condensate versus temperature}
The first important problem
to study is how the order parameter changes with the
temperature for this holographic superconductor with
fixed $A_x^{(0)}$. From now on we suppose $A_x^{(0)}$ is the
supercurrent $j_x$ and $A_x^{(1)}$ is the superfluid velocity $-v_x$.
From Fig. 3 it can be seen that when the supercurrent is not zero, there are two solutions of the order parameter corresponding to a fixed temperature.  We also show that the solution with lower value of the order parameter takes
a larger free energy than the solution with larger values of the order parameter and therefore
it is unfavorable. In Fig. 4 we present the free energy of a fixed current $j_x=1/100$ for the two branches of
solution. It can be obviously seen that the solution with a larger value of the order parameter has a lower free energy.
The critical temperature decreases when
the current increases, which indicates that there should exist a critical current
above which there is no superconductivity. When one
lowers the temperature from above the critical temperature, the order of phase transition
at the critical temperature for a fixed current should be of first order, since the order parameter jumps from zero to a finite value at the critical temperature. Such a jump will certainly change the energy and so requires some latent heat, which
implies that the phase transition should be of first order. This conclusion is the same as the one we shall give by observing the curve of the current $j_x$ versus the superfluid velocity $v_x$ at a fixed temperature soon.
For the $AdS_4$ $s$-wave holographic superconductor with current,
the order parameter is also bivalued, and the states with lower value of the condensate have a larger free energy than
their counterparts with larger values of the condensate at the same temperature.\cite{20}

\begin{figure}
\label{fig:zeroKSecondSound}
\includegraphics[width=8cm,clip]{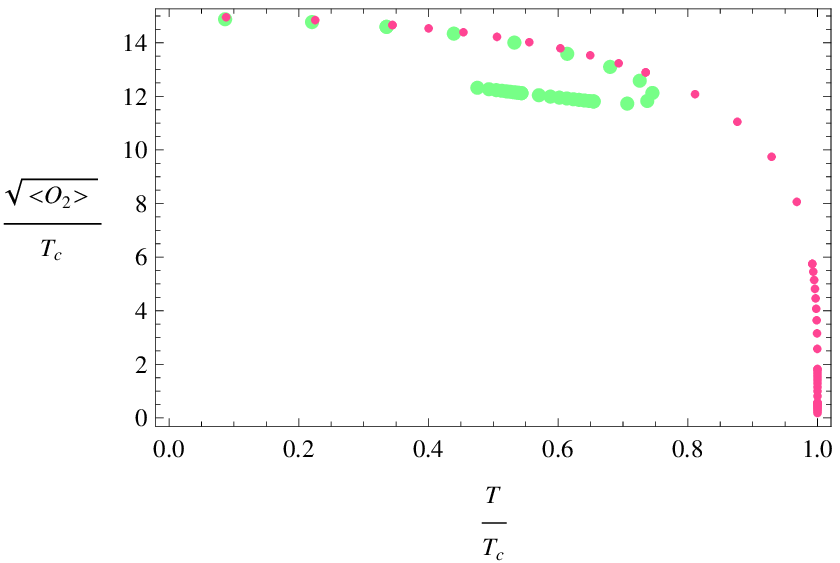}
\includegraphics[width=8cm,clip]{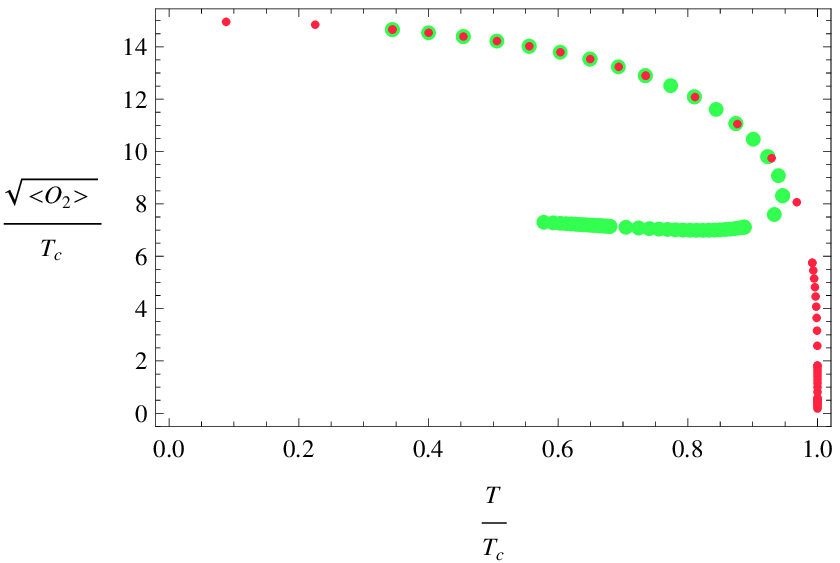}
\includegraphics[width=8cm,clip]{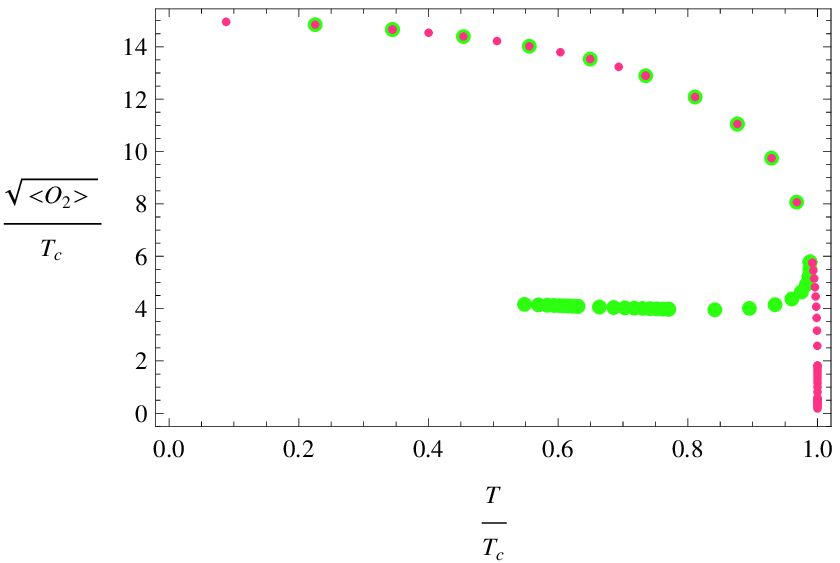}
\includegraphics[width=8cm,clip]{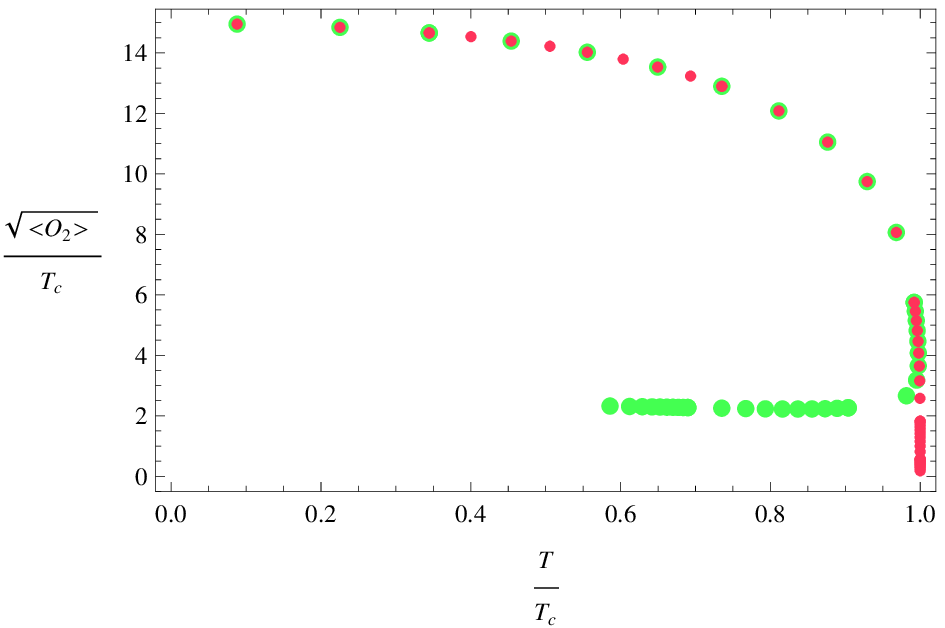}
\caption{Plot of the order parameter versus the temperature for different fixed values of
current(The $A_x^{(0)}$). We show three different plots of
order parameter versus the temperature for $j_x$=1/10, 1/100, 1/1000, 1/10000 (from the left to the right).
In which we also plot the order parameter versus temperature when $A_x=0$ for
compare.
We have identified $A_x^{(0)}$ with $j_x$ already in the plots.(
The green plots are the condensate with fixed $A_x^{(0)}$, the red plots represent the
condensate with $A_x=0$.) }
\end{figure}

\begin{figure}
\includegraphics[width=7.5cm,clip]{ax02.eps}
\includegraphics[width=7.5cm,clip]{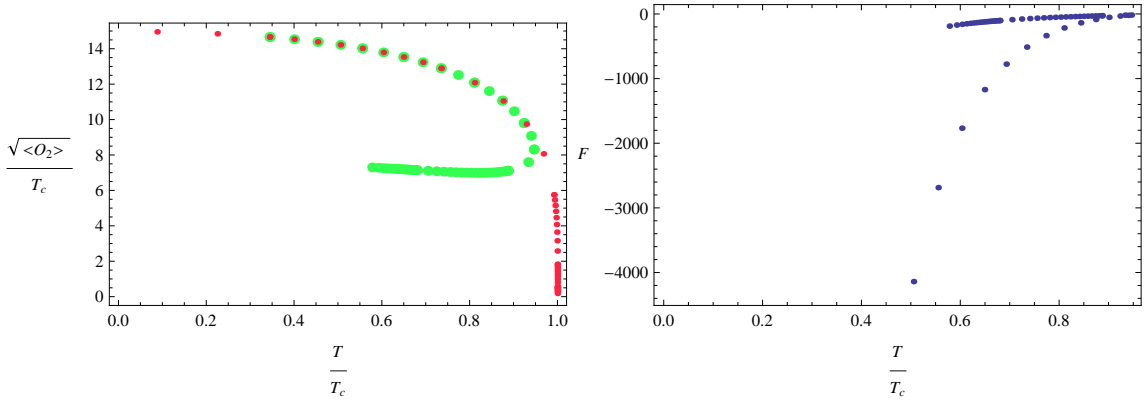}
\caption{The free energy of the two branches of solution when $j_x=1/100$,
the upper dotted line corresponds to the
points with a lower value of the condensate at a given temperature. The left panel shows the corresponding plot of the condensation
versus $T$. It can be seen that the lower branch solution corresponds indeed to states with larger free energy and is thus metastable. \label{fig:condensation} }
\end{figure}

\subsection{Results that agree with the G-L theory}
In this section we give more evidences that $A_x^{(0)}$ behaves
the same as a supercurrent in a superconductor.
These results indicate that we should interpret
$A_x^{(0)}$ as the supercurrent $j_x$ in the boundary theory.

\subsubsection{Current via the superfluid velocity}
Another physical property by which one can compare the difference between the gravity model of superconductor and the G-L theory is the relation between the current and the superfluidity velocity at a fixed temperature. From this relation we can also get the information of the
phase transition at the critical current.

\begin{figure}
\includegraphics[width=7.6cm,clip]{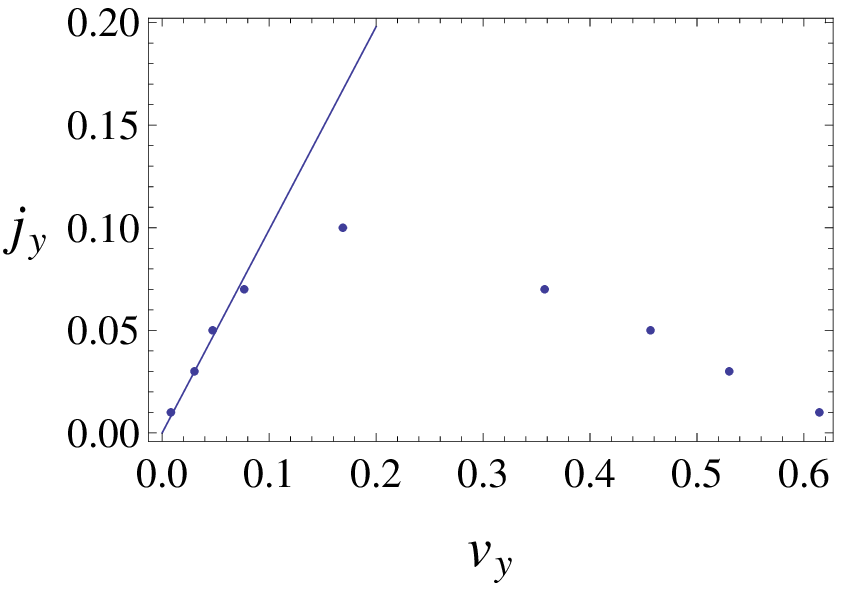}
\includegraphics[width=8cm,clip]{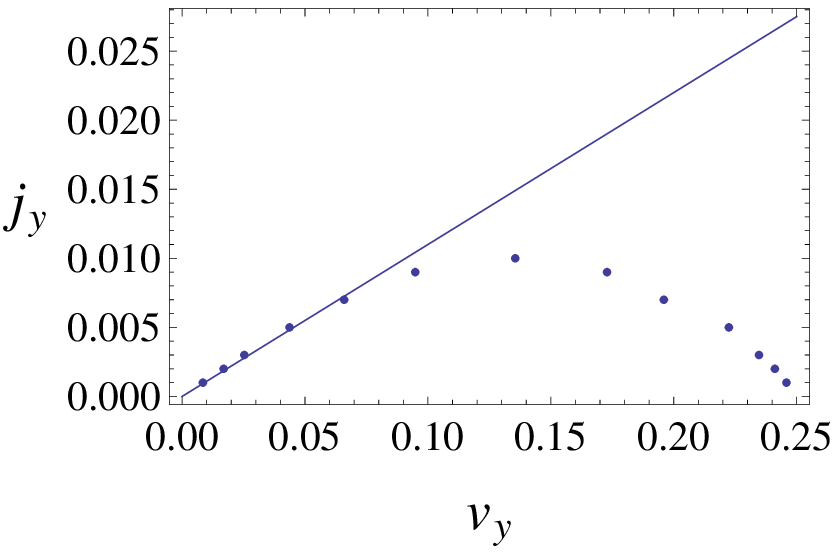}
\caption{Plots of the current $j_x$ versus the superfluid velocity $v_x$ at a fixed temperature.
The two panels above correspond to $T/T_c=0.745385, 0.946225$ (from left to right),
at which the critical currents $j_c$ are $1/10$ and $1/100$, respectively.
It is clear that the $v_x$ and $j_x$ have a linear relationship when $j_x$ is not very large.
 \label{fig:condensation2} }
\end{figure}

The two plots in Fig. 5 correspond to the two temperatures $T/T_c=0.745385 $ and $ 0.946225$. For these two temperatures the critical currents are $j_c=1/10$ and $j_c=1/100$,
respectively. It can be clearly seen that at the temperatures close to $T_c$,
where the G-L theory works very well, the plots of $j_x$ versus $v_x$ are the same
as that of G-L theory,
in which the $j_x$ and $v_x$ have a linear relationship when $j_x$ is much smaller than $j_c$.
From these plots we can also infer the order of the phase transitions
at critical current or critical velocity. For the two plots, the maximal velocity corresponds to a vanishing
current, which means that the phase transitions at critical velocities the $v_c=0.65$ and $v_c=0.25$ are of
second order. However, the maximal current corresponds to a non-vanishing
velocity, which means that the phase transitions at critical currents $j_c=1/10$ and $j_c=1/100$ are of
first order.

\subsubsection{The critical current via temperature}
In this subsection we study the critical current $j_c$
for different $T$ near $T_c$ to compare the results with those of
the G-L theory. As predicted by G-L theory, $j_c$ is proportional to $(T_c-
T)^{3/2}$ when the temperature is close to $T_c$. As illustrated in Fig. 6, this scaling behavior is indeed obeyed by holographic
superconductors for temperatures close to $T_c$, which is also the case in the $(2+1)$ dimensional $s$-wave model\cite{20}. Another prediction of the G-L theory is that, at any fixed temperature, the
norm of the condensate decreases monotonically with respect to the velocity from
its maximal value. And the maximal value $\sqrt {\langle O_2 \rangle}_\infty$ corresponds to zero velocity and zero current. As shown in Fig. 5, the critical current is reached before the velocity reaches its maximal value. The norm of the condensate has an intermediate value $\sqrt{ \langle  O_2 \rangle}_c$ at the maximal current.
The G-L theory tells us that the ratio of $\sqrt {\langle O_2 \rangle}_c$ to  $\sqrt{ \langle O_2 \rangle}_\infty$
is exactly $2/3$. From Fig. 7 it can be seen that this is indeed the case for the $(1+1)$ dimensional $s$-wave holographic superconductor.

\begin{figure}
\includegraphics[width=8cm,clip]{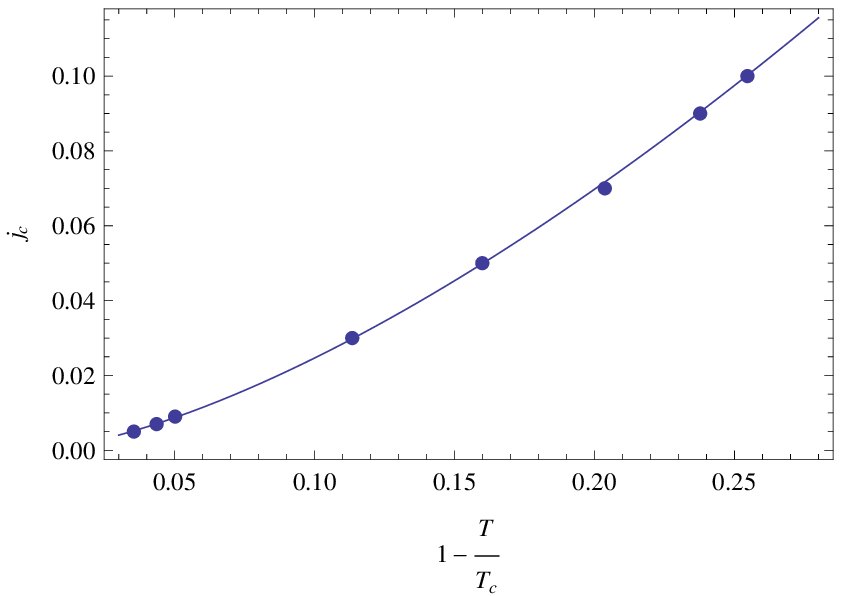}
\includegraphics[width=8cm,clip]{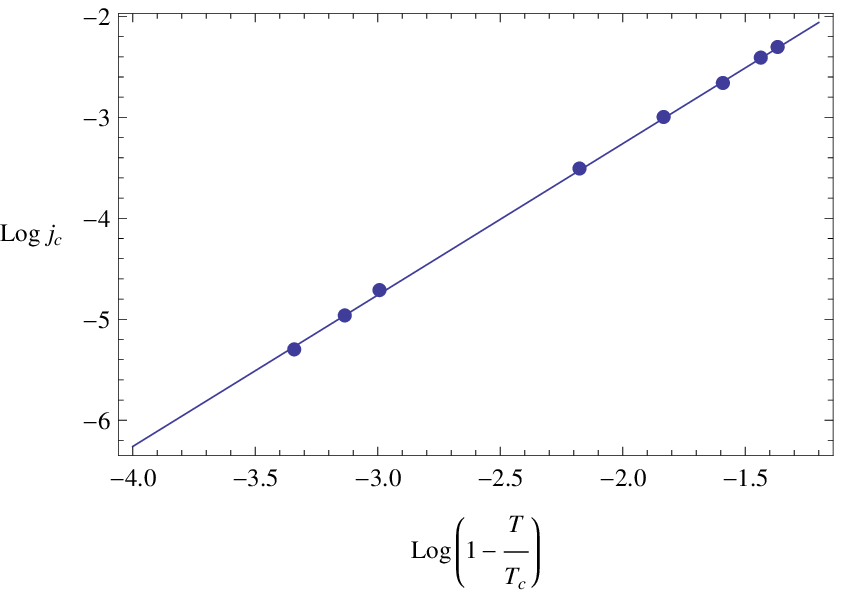}
\caption{Plot of the critical current versus the temperature. The right panel shows a
log-log plot from which we can read off the critical exponent, getting 1.499, which agrees
with the expected G-L scaling of 3/2 within numerical precision. The left panel shows
the $j_c$ versus $(1-T/T_c)$, the solid line is $0.78(1-T/T_c)^{3/2}$.. \label{fig:condensation2} }
\end{figure}

\begin{figure}
\begin{center}
\includegraphics[width=10cm,clip]{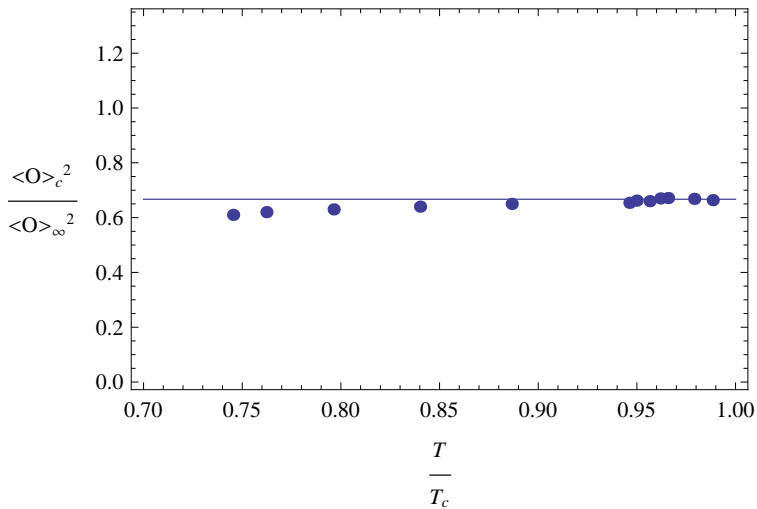}
\caption{Plot of the ratio ($\sqrt{\langle O_2 \rangle}_c / \sqrt{\langle O_2 \rangle}_\infty)^2$ versus the temperature. The solid line
corresponds to the value of 2/3 predicted by the G-L theory and it also appears in the $AdS_4$ $s$-wave model. \label{fig:condensation} }
\end{center}
\end{figure}

\section{The superconductor with fixed $A_x^{(1)}$}
Before we give the results when we solve the
EOMs with fixed $A_x^{(1)}$.
We would like to emphasis that there is only one kind of quantization of the theory,
we can only take $A_x^{(0)}$ is the source.
However, fixing $A_x^{(1)}$ does not contradict
the statement in \cite{21}, we are simply in another ensemble
which can be changed to the ensemble in last section with fixed $A_x^{(1)}$
by a Legendre transformation\cite{22}.
From the numerical computation we find that to solve the
EOMs with fixed $A_x^{(0)}$ is indeed much more difficult
numerically, it will take very long time for the
shooting method to give a solution.
In Fig. 7 we give two samples of solutions for
fixed $A_x^{(1)}= -0.1$ and $-0.5 $ respectively.
It can be seen that the phase transition with the small
values of velocity is of second order, which confirms
our discussion in section 3.2.1.

\begin{figure}
\includegraphics[width=8cm,clip]{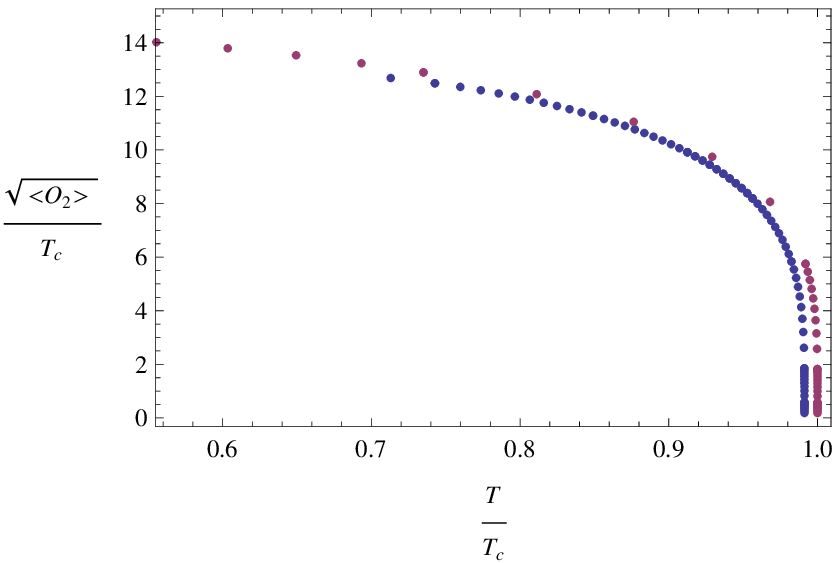}
\includegraphics[width=8cm,clip]{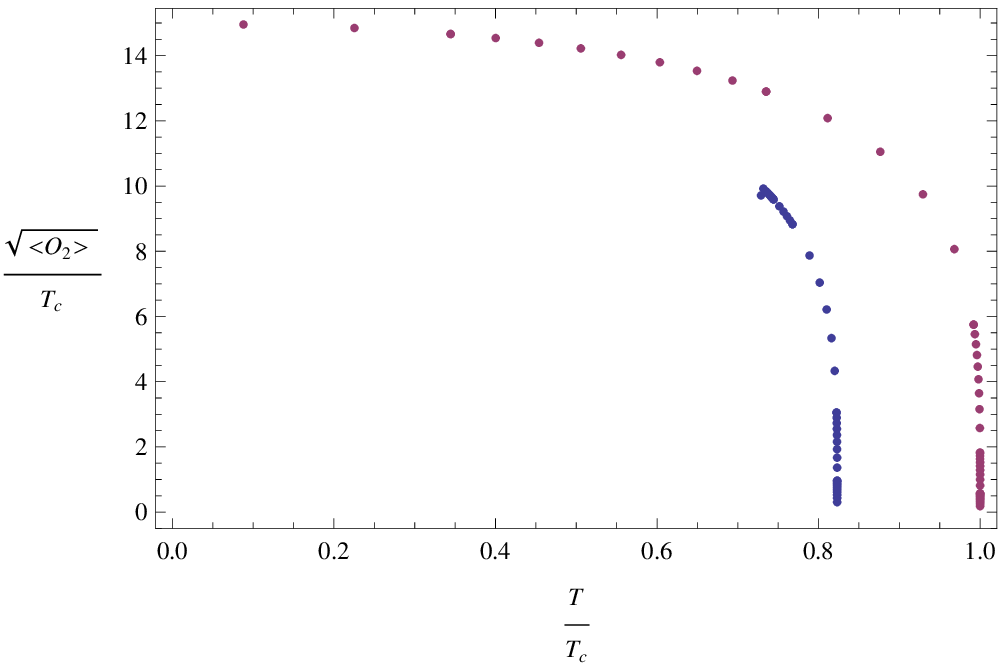}
\caption{Plot of the order parameter versus the temperature for different fixed values of
velocity (The $-A_x^{(1)}$). We show three different plots of
order parameter versus the temperature for $-A_x^{(1)}$=1/10, 1/2 (from the left to the right).
It can be seen that the phase transition is of second order.(
The dark blue plots are the condensate with fixed $A_x^{(0)}$, the dark red plots represent the
condensate without $A_x$.)
\label{fig:condensation2} }
\end{figure}

\section{Conclusion}

With the fact the Maxwell gauge fields
$A_t$ and $A_x$ behave
as in Eq. (2.8) and Eq. (2.9) on the $AdS_3$ boundary, and an important question arises as how to interpret
the coefficients $A_i^{(0)}$ and $A_i^{(1)}$.
Formally, this question was already answered in \cite{18,21,22},
the source term  $A_t^{(0)}$ is the charge density and
$A_t^{(0)}$ is the chemical potential.
In this paper we have confirm the conclusion given in \cite{18,21,22} by
studying the properties of the $(1+1)$ dimensional
holographic superconductor with nonzero $A_x$.
By solving the EOMs we found that if we interpret $A_x^{(0)}$ and $-A_x^{(1)}$ in
Eq. (2.9) as the supercurrent $j_x$ and the superfluidity
velocity $v_x$ respectively, the results near the
critical temperature agree qualitatively with several
properties of the Ginzburg-Landau theory:
\begin{itemize}
\item The phase transition with supercurrent is of first order
\item The critical supercurrent $j_c$ is proportional to $((T_c-T)^{3/2})$
\item The relation between velocity and supercurrent is linear almost all the way up to a given
maximum velocity
\item The squared ratio
of the maximal condensate to the minimal condensate is equal to two thirds
\end{itemize}
Furthermore, the $\Phi^{(0)}$ in Eq. (2.8) is the charge density
and $\Phi^{(1)}$ is the chemical potential. We summarize
these results as
\begin{equation}
\Phi=\mu -\rho \ln z,
\end{equation}
and
\begin{equation}
A_x=-v_x + j_x \ln z
\end{equation}
on the boundary,
where $\mu$ is the chemical potential, $\rho$ is the charge density,
$v_x$ is the superfluid velocity and $j_x$ is the supercurrent.

We state that there should be a supercurrent
in the $(1+1)$ dimensional holographic superconductor when we include $A_x$ in the bulk theory, which corresponds to the coefficient of the
logarithmic term on the boundary (see Eq. (2.9)).
The existence of supercurrent in a holographic superconductor
means there should also be superconducting
tunneling effect by building a holographic Josephson Junction.\cite{27,28,29,30}
Then another way to confirm the
conclusion that $A_x^{(0)}$ is the supercurrent is to
find the holographic Josephson Junction
in the $AdS_3$ setup of superconductors.
We leave this for future work.
\section*{Acknowledgements}
It is a pleasure to thank Xin Gao, Matthias Kaminski, Hai-Qing Zhang and Zhe-Yong Fan for valuable discussions.
H. B. Zeng is supported by the Fundamental Research Funds for the Central Universities (Grant No. 1107020117) and
the China Postdoctoral Science Foundation (Grant No. 20100481120).

\end{document}